\documentclass{iaus}
\usepackage{graphicx}

\title[SuperLupus] 
{SuperLupus: A Deep, Long Duration Transit Survey}

\author[Daniel D. R. Bayliss, Penny D. Sackett \& David T. F. Weldrake]   
{Daniel D. R. Bayliss$^1$,
Penny D. Sackett$^1$, \and
David T. F. Weldrake$^2$}

\affiliation{
$^1$Research School of Astronomy and Astrophysics, The
  Australian National University (ANU), \\Mount Stromlo Observatory, Cotter
  Road, Weston Creek, ACT 2611, Australia \\ email: {\tt
    daniel@mso.anu.edu.au} \\ email: {\tt psackett@mso.anu.edu.au} \\[\affilskip]
$^2$Harvard Smithsonian Center for Astrophysics, 60 Garden Street
  MS-51, \\Cambridge, MA 02138, USA \\email: {\tt dweldrak@cfa.harvard.edu}}

\pubyear{2008}
\volume{253}  
\pagerange{119--126}
\jname{Transiting Planets}
\editors{F. Pont, ed.}
\begin{document}

\maketitle

\begin{abstract}
SuperLupus is a deep transit survey monitoring a Galactic
Plane field in the Southern hemisphere.  The project is building on
the successful Lupus Survey, and will double the number of images of
the field from 1700 to 3400, making it one of the longest duration
deep transit surveys.  The immediate motivation for this expansion is
to search for longer period transiting planets (5-8 days) and smaller radii
planets.  It will also provide near complete recovery for the shorter
period planets (1-3 days).  In March, April, and May 2008 we obtained
the new images and work is currently in progress reducing these new data.

\end{abstract}

\firstsection 
\section{The Original Lupus Survey}
\begin{figure}[!ht]
 \vspace*{-2.5 cm}
\begin{center}
 \includegraphics[width=8.5cm]{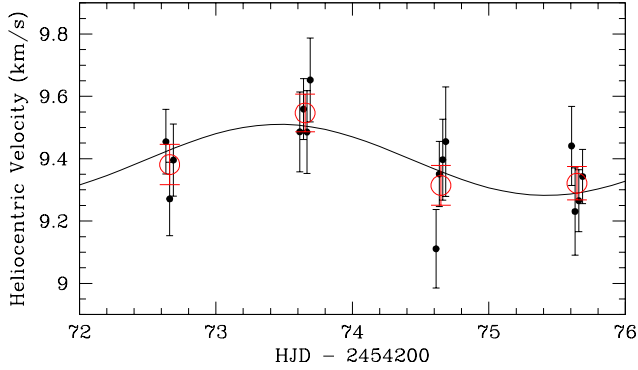} 
 \vspace*{-0.5 cm}
 \caption{Radial velocity measurements for Lupus-TR-3 from MIKE on Magellan II (Clay) (\cite[Weldrake et
  al. (2008)]{Weldrake08b}).  Small solid circles are the individual
  radial velocity measurements, with the error bars determined from the actual scatter in the
  orders. Large open circles are uncertainty-weighted nightly
  averages.  The solid line is the best-fit sinusoid with only the
  period fixed (from the photometry).  The fitted phase matched the
  photometrically-determined phase to within 0.13~days, well within
  fit uncertainties.}
   \label{lupus3}
\end{center}
\end{figure}

The Lupus Survey was a deep transit survey of a 0.66 square degree
patch of sky near the Galactic Plane (b=11$^{\circ}$).  The survey was
conducted using the ANU 40 Inch Telescope at Siding Spring
Observatory, Australia in May and June of 2005 and 2006.  In
total, 1783 good quality images of the field were obtained.  The images were 5-minute
exposures taken in a wide V+R filter.  Time series photometry was performed for
110,372 stars in the field, with 16,134 of those stars having a
precision of $\sigma<0.025$.  The transiting planet candidates and results from this
survey will be published in \cite[Bayliss et al. (2008)]{Bayliss08}.
The discovery of a Hot Jupiter in the field, Lupus-TR-3b, has been
published in a separate letter (\cite[Weldrake et
  al. (2008)]{Weldrake08b}), and the radial velocity follow-up
confirming the discovery is set out in~Figure~\ref{lupus3}. 
Additionally, 494 new variables were discovered in the survey field.  These variables have
been cataloged in \cite[Weldrake \& Bayliss (2008)]{Weldrake08}.  

\section{SuperLupus: Extending the Survey Duration}
The duration of a transit survey is critical to its prospects of
success, and underestimating required durations may
have contributed to early surveys not discovering transiting planets in the
numbers expected (\cite[Pont, Zucker, \& Queloz (2006)]{pont06}).

\begin{figure}[!ht]
\begin{center}
 \includegraphics[width=11.0cm]{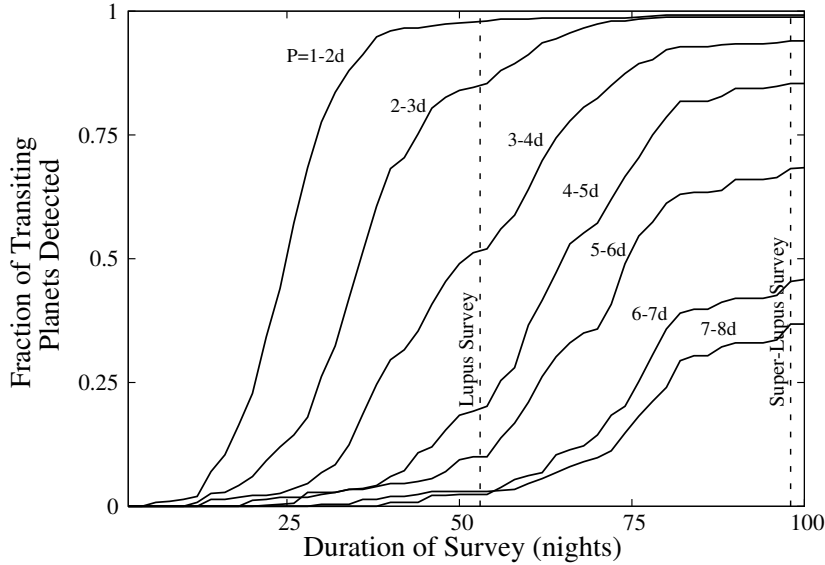} 
 \caption{The fraction of planets detected as a function of the duration
 of the survey for 7 different period bins.  These simulations show the benefit of
 moving from the Lupus Survey (50 nights; left dashed line)
 to the expanded SuperLupus Survey (100 nights; right dashed line),
 both in terms of completeness at
 shorter periods and greatly increased sensitivity at longer periods.}
   \label{recover}
\end{center}
\end{figure}

We  modeled the effect of increasing our survey duration by looking
at the transit recoverability
for Hot Jupiters as a function of planetary period and the duration of
the survey.  We created 1000 transiting Hot
Jupiters in each of 7 period bins, ranging from \mbox{2-3~days} to 7-8~days.
Each transit lightcurve was given a random phase.  We then convolved these transits with a window function
out to 100 nights based on actual weather logs taken from the Siding
Spring Observatory site.  A detection was equated with observing the
equivalent of three full transits.  The results of the simulation are
plotted \mbox{in Figure~\ref{recover}}.  These results indicate that by 100
nights (the SuperLupus duration) we will detect nearly all transiting Hot
Jupiters in the field with periods from 1 to 3 days.  The fraction of
longer period Hot Jupiters we can detect will rise significantly from
the original Lupus survey, especially planets with 5-8 day periods, to
which the original survey was very insensitive.  Our new
dataset should also allow us to increase our
sensitivity to smaller radius planets, as more data-points in the
lightcurve will increase our S/N.

\section{SuperLupus: The New Data}
Based on these simulations, we have initiated the SuperLupus project to
expand the original Lupus
Transit Survey by imaging the field again in 2008.  The instrument
set-up and observational strategy is identical to that used in the
original survey (see Table \ref{survey}).

Data has now been taken in the months of March, April and May 2008.  We
have approximately 2500 new images, and expect at least 70\% of
these will be of sufficient quality to use in the production of high
precision time series photometry.

\section{Aperture Photometry: Source Extractor} 
The photometry for the original survey was produced using Difference
Imaging Analysis (DIA: \cite[Alard \& Lupton (1998)]{alard98},
\cite[Wozniak (2000)]{wozniak00}).  This method is well suited to crowded
fields.  The Lupus field is \textit{moderately} crowded, being only
11$^{\circ}$ above the Galactic Plane.  In order to test how aperture
photometry will compare to the DIA photometry in this r\'egime, we
used the IRAF package DAOPHOT and the
Source Extractor software (\cite[Bertin \& Arnouts (1996)]{bertin96}) to produce time
series photometry on a small subset of the 2006 images.  These tests
indicated that Source Extractor, with its more sophisticated background
subtraction, gave slightly better results than DAOPHOT for our images, and that it
compared well to DIA photometry.  Source Extractor is also a very fast
algorithm, and this will  allow us to perform photometry using multiple apertures and
select the one best suited to the star and its environment.

\begin{table}[!h]
  \begin{center}
  \caption{Properties of the SuperLupus Transit Survey}
  \label{survey}
  \vspace{2mm}
  \begin{tabular}{ll}\hline\hline
     &  \\
    Telescope        & ANU 40 Inch Telescope (1.0m aperture)               \\
    Site             & Siding Spring Observatory, Australia    \\
                     & Lat.:  $-$31$^{\circ}$16$^{\rm{m}}$36$^{\rm{s}}$\\
                     & Long.:
    $-$9$^{\rm{h}}$56$^{\rm{m}}$16$^{\rm{s}}$ W\\

    Field of View    & 0.66 sq degrees    \\ 
    Cadence          & 6 minutes          \\
    Field Location   & Lupus (b=11$^{\circ}$)\\
                     & RA:15$^{\rm{h}}$30$^{\rm{m}}$36.3$^{\rm{s}}$,\\
                     & Dec:$-$42$^{\circ}$53$'$53.0$''$ (J2000)\\
    Pixel Size       & 15 microns, 0.375 pixels/arcsecond\\
    Filter           & Custom V+R filter\\
    Stars Monitored  & 110,372, with 16,134 to $\sigma<0.025$ mags\\
    Number of Images & 1783, expanding to $\approx$3400 with SuperLupus\\
  & \\\hline
  \end{tabular}
 \end{center}
\end{table}

\end{document}